\documentclass[namedreferences]{SolarPhysics}
\usepackage{epsfig}

\newcommand{\be}{\begin{equation}}
\newcommand{\ee}{\end{equation}}
\newcommand{\beq}{\begin{eqnarray}}
\newcommand{\eeq}{\end{eqnarray}}
\renewcommand*{\tablename}{~~~~~~~}

\begin{document}
\begin{article}
\begin{opening}

\title{AN ASYMMETRIC CONE MODEL FOR HALO CORONAL MASS EJECTIONS}

\author{G. \surname{MICHALEK}}
\runningauthor{MICHALEK} \runningtitle{An Asymmetric Cone Model }


\institute{
              Astronomical Observatory of Jagiellonian University, Cracow, Poland\\
                          ({e}-mail: michalek@oa.uj.edu.pl)}

\date{Received: date}
% The correct date will be entered by the editor

%\maketitle

\begin{abstract}
Due to projection effects,  coronagraphic observations
 cannot uniquely  determine parameters relevant to the geoeffectiveness of CMEs, such  as the
 true propagation
speed, width, or source location. The Cone Model for Coronal Mass
Ejections (CMEs) has been studied in this respect  and it could be
used to obtain these parameters. There are evidences that some CMEs
initiate from a flux-rope topology. It seems that these CMEs should
be elongated along the flux-rope axis and the cross section of the
cone base should be rather elliptical than circular. In the present
paper we applied an asymmetric cone model to get the real space
parameters of frontsided halo CMEs (HCMEs) recorded by SOHO/LASCO
coronagraphs in 2002. The cone model parameters are generated
through a fitting procedure to the projected speeds measured at
different position angles on the plane of the sky. We consider
models with the apex of the cone located at the center and  surface
of the Sun. The results are compared to the standard symmetric cone
model.
\end{abstract}

%\keywords{Coronal mass ejections---solar physics---space weather}

\end{opening}
\section{Introduction}
A halo coronal mass ejection (HCME) was first recorded by Howard in
1982 (Howard \emph{et al.}, 1982). Since then, HCME are routinely
recorded in white light by coronagraphs placed in  space. In
coronagraphic observations, HCMEs appear as an enhancement
surrounding the entire occulting disk.  HCMEs originating close to
the disk center are often responsible for the severest geomagnetic
storms (Gosling, 1993; Kahler, 1992; Webb \emph{et al.}, 2000). For
space weather forecast it is very important to determine the kinetic
and geometric parameters describing HCMEs. Unfortunately
coronagraphic observations are subjected to projection effects.
Viewing in the plane of the sky does not allow to determine the true
3-D space velocity, width and source location of a given CME. It is
widely accepted that the geometrical structure  of CMEs may be
described by the cone model (e.g., Howard \emph{et al.}, 1982;
Fisher and Munro, 1984; St.Cyr \emph{et al.}, 2000; Webb \emph{et
al}, 2000; Zhao \emph{et al.}, 2002; Michalek \emph{et al.}, 2003;
Xie \emph{et al.}, 2004; Xue \emph{et al.}, 2005). Assuming that the
shape of HCMEs is a cone and they propagate with constant angular
widths and speeds, at least in their early phase of propagation, a
technique was developed (Michalek \emph{et al.}, 2003) which can
determine the following parameters: the linear distance $r$ of the
source location measured from the solar disk center,  the angular
distance $\gamma$ of the source location measured from the plane of
the sky, the angular width $\alpha$ (cone angle =$0.5\alpha$) and
the 3-D space velocity $V_{S}$ of a given HCME. This technique
required measurements of the sky-plane speed and the moment of the
first appearance of the halo CME above opposite limbs. If we
determine spatial parameters using only two measurements, large
random errors would occur. What is more, this technique was limited
to asymmetric events not originating from close to the center of the
Sun. A similar cone model was used recently by Xie
\emph{et~al.}~(2004) to determine the angular width and orientation
of HCMEs. To improve accuracy, in the present attempt the space
parameters of HCMEs are determined by fitting the cone model to
projected speeds ($V_{P}$) obtained from height-time plots at
different position angles. Although  many thousands of CMEs were
recorded by LASCO coronagraphs the 3D structure of  CMEs is still
open question. Many authors believe that CMEs originate from a
flux-rope geometry, (e.g., Chen \emph{et al.}, 1997; Dere \emph{et
al.}, 1999; Chen \emph{et al.}, 2000; Plunket \emph{et al.}, 2000;
Forbes 2000; Krall \emph{et al.}, 2001; Chen and Krall 2003). If
CMEs have a flux-rope geometry,  they should be elongated along the
flux-rope axis and the cross section of the cone base should be
rather elliptical than circular. In the present approach we consider
the asymmetric cone model where an eccentricity and orientation of
the cone base are new free parameters. We try  to identify where to
locate the apex of the cone, either at   the center of the Sun (Zhao
\emph{et al.}, 2002; Xie \emph{et al.}, 2004) or on the solar
surface (Michalek \emph{et al.}, 2003). It is important to note that
the real elliptical cone model was first developed by Cremades and
Bothmer (2004). The model was introduced based on observations of
cylindrical shaped CMEs (Cremades and Bothmer, 2004, 2005). They
applied it to 32 halo CMEs. In their approach, the best parameter
values describing the ellipse are determined from a LASCO image
sequence that showed a sharp leading edge. In
 our method we derive the best-fit parameter values for  halo
CMEs by working in the velocity space.  The paper is organized as
follows: in Section~2 the asymmetric cone model is presented,
numerical simulations and fitting procedure are explained in
Section~3 and in Section~4 the results are described. Final
conclusions are given in Section 5.

\section{The Asymmetric  Cone Model of CMEs}
In the projection on the sky, most of the CMEs (especially limb
events)
 observed by the LASCO coronagraphs look like cone-shaped blobs.   The
observed angular widths, for many limb events, remain nearly
constant as a function of height (see, e.g., Webb \emph{et al.},
1997). Most of them propagate with a constant radial frontal speed
but many slow CMEs gradually accelerate whereas many fast CMEs
decelerate (St. Cyr \emph{et al.}, 2000; Sheeley \emph{et al.},
1999; Gopalswamy \emph{et al.}, 2001; Yashiro \emph{et al.}, 2004).
Assuming that CMEs propagate with a constant velocity and angular
width, many authors reproduced them (Howard \emph{et al.}, 1982;
Fisher and Munro 1984; Zhao \emph{et al.}, 2002; Michalek \emph{et
al.}, 2003, Xie \emph{et al.}, 2004) using the cone model defined by
four parameters: velocity, angular width, and orientation of the
central axis of the CME (longitude and latitude). As was mentioned
above, many CMEs originate from the flux-rope geometry and their
cone shape may not be perfectly symmetrical. This was also
demonstrated recently by Moran and Davila (2004), Cramades and
Bothmer (2004) and Jackson \emph{et al.} (2004). This encouraged us
to introduce the asymmetric cone model.
 In the  asymmetric cone model we  assume that (1) the shape of CMEs is a cone but its
cross section is not a circle but ellipse and (2) the velocity and
shape of CMEs (angular widths measured along the major and minor
axes of the ellipse) remain constant during  their early phase of
propagation. The eccentricity and orientation of the elliptic cone
cross section appear as  the additional two parameters of the model.
We also consider  two possibilities, namely the apex of the cone is
located at the center of the Sun and on the surface of the Sun.

 To obtain relationships between the measured  velocities and the cone model parameters we had to
 apply the transformation between two coordinate systems: First, a heliocentric coordinate system (HCS) ($x_{h},
y_{h}, z_{h}$), where $x_{h}$ points to Earth, $z_{h}$ points north
and  $ y_{h}-z_{h}$ defines the sky plane. Second, an apex-centered
cone coordinate system (CCS) ($x_{c}, y_{c}, z_{c}$) whose origin is
at  the apex of the cone,  $z_{c}$ is the cone axis, and $x_{c}-
y_{c}$ defines the plane parallel to the base of the cone. The
orientation of the cone is determined by heliographic longitude
($\varphi$) and latitude ($\lambda$) measured from the central
meridian and the solar equator, respectively. If the apex of the
cone is at the center of the Sun, the transformation from the HCS to
the CCS can be carried out by double rotations. The first rotation
is about the $x_{h}$ axis through the angle $\varphi$. The second
rotation is about the $y_{h}$ axis through the angle
$90^o-\lambda=\vartheta$. This transformation in matrix notation can
be written as:
$$ \left(\matrix{x_{h} \cr y_{h} \cr z_{h}}\right)=
\left( \matrix{& \cos\vartheta \cos\varphi & -\sin\varphi
& \cos\varphi \sin\varphi \cr &\cos\vartheta \sin\vartheta&
\cos\varphi & \sin\varphi\sin\vartheta \cr  & -\sin\varphi & 0 & \cos\vartheta }\right) \left(\matrix{x_{c} \cr
y_{c} \cr z_{c }}\right)    \eqno(1)$$ or

$$ \left(\matrix{x_{c} \cr y_{c} \cr z_{c}}\right)=\left( \matrix{& \cos\vartheta \cos\varphi & \cos\vartheta\sin\varphi &- \sin\vartheta \cr &- \sin\varphi&
\cos\varphi & 0 \cr  & \cos\varphi\sin\vartheta &
\sin\varphi\sin\vartheta & \cos\vartheta }\right)
\left(\matrix{x_{h} \cr y_{h} \cr z_{h }}\right).  \eqno(2)$$ In
Figure~1, the cone model topology and the relationship between the
HCS and the CCS is shown. If the apex of the cone is on the surface
of the Sun, these relationships are slightly modified. The
transformation from the HCS to the CCS can be carried out by double
rotations and additional shifting of the origin to the surface of
the Sun. Then we obtain: {\footnotesize

$$ \left(\matrix{x_{h} \cr y_{h} \cr z_{h}}\right)=\left( \matrix{& \cos\vartheta \cos\varphi & -\sin\varphi & \cos\varphi \sin\varphi \cr &\cos\vartheta \sin\vartheta&
\cos\varphi & \sin\varphi\sin\vartheta \cr  & -\sin\varphi & 0 &
\cos\vartheta }\right) \left(\matrix{x_{c} \cr y_{c} \cr z_{c
}}\right)+
\\
\left(\matrix{R_{S}\sin\vartheta \cos\varphi \cr R_{S}\sin\vartheta
\sin\varphi \cr R_{S}\cos\vartheta}\right) \eqno(3)$$ } or
{\footnotesize
$$ \left(\matrix{x_{c} \cr y_{c} \cr z_{c}}\right)=\left( \matrix{& \cos\vartheta \cos\varphi & \cos\vartheta\sin\varphi &- \sin\vartheta \cr &- \sin\varphi&
\cos\varphi & 0 \cr  & \cos\varphi\sin\vartheta &
\sin\varphi\sin\vartheta & \cos\vartheta }\right)
\left(\matrix{x_{h} \cr y_{h} \cr z_{h
}}\right)-\left(\matrix{R_{S}\sin\vartheta \cos\varphi \cr
R_{S}\sin\vartheta \sin\varphi \cr R_{S}\cos\vartheta}\right)
\eqno(4)$$ } where $R_S$ is the solar radius. The topology of the
cone model in this case is demonstrated in Figure~2. Using these
transformations, we can find the angle ($\sigma$) between a
generatrix of the cone and the plane of the sky (Xue \emph{et al.},
2005):
$$\sin\sigma={\cos{\alpha\over 2}\cos\varphi\sin\vartheta-A\sqrt{\cos^{2}\varphi\sin^{2}\varphi+A^{2}-\cos^{2}{\alpha\over
2}}\over \cos^{2}\varphi\sin^{2}\varphi-A^{2}} \eqno(5) $$ where
$$ A=\cos \psi \sin\varphi\sin\vartheta+\sin \psi\cos\vartheta  \eqno(6) $$
and $\psi$ is the azimuthal angle of a given cone generatrix in the
plane of sky. The basic equation for our consideration is the
relationship  between the projected velocities ($V_{P}$, derived
from LASCO height-time plots) and the 3-D space velocities ($V_{S}$,
a parameter in  the cone model):
$$V_{P}=V_{S}\cos\sigma.  \eqno(7)$$
We have to note that we presented  the relationships of Equations
(6) and (7) only for the model in which  the cone apex is at the
center of the Sun. For  another possibility (when the cone apex is
placed on the surface of the Sun), equations become more
complicated. It is evident that $V_{P}$ depends on $V_{S}$, the
location of the origin (cone apex) on the Sun (longitude-$\varphi$,
latitude-$\lambda$), and the width ($\alpha$) of the cone. The above
formulae are also applicable to  the symmetric cone model. We
improved these formulae by considering an asymmetric cone namely the
base of the cone has an elliptical shape. In Figure~3 the topology
of the cone base in the cone coordinate system is presented. This
modification introduces  the eccentricity and orientation of the
ellipse as new parameters. In this paper the eccentricity is defined
as:
$$e=\sqrt{\alpha_{\textrm{\tiny{MAX}}}^{2}-\alpha_{\textrm{\tiny{MIN}}}^{2} \over \alpha_{\textrm{\tiny{MAX}}}^{2}}  \eqno(8)$$
where $\alpha_{\textrm{\tiny{MAX}}}$ and
$\alpha_{\textrm{\tiny{MIN}}}$ are the angular widths of the cone
cross section along the major and minor axes  of the ellipse,
respectively. This parameter should depend on the separation of the
foot points and
 width of the flux rope. The orientation of the ellipse (the position angle of the major axis) should   depend on
the orientation of the foot points of the flux rope. The flux-rope
could be oriented in a random way on the solar disk, so we need to
consider the next free parameter of the model: the orientation of
the ellipse ($\beta$) which is defined as the azimuthal angle  of
the main axis of the ellipse measured in the plane $x_c-y_c$. To use
the formula (7) to determine the space parameters of CMEs we need to
express $\alpha$ as a function of position angle ($\psi$) measured
in the plane of the sky. First, we can define $\alpha$ as a function
of  angle $\gamma$. From the definition of the ellipse and  Figure 3
we can write:
 $$\alpha_{1/2}=\sqrt{({\alpha_{\textrm{\tiny{MAX}}}\over 2})^{2}\cos(\delta-\beta)+({\alpha_{\textrm{\tiny{MIN}}}\over 2})^{2}\sin(\delta-\beta)}.
 \eqno(9) $$
 This equation shows the dependence of $\alpha$ on the position angle
 ($\gamma$), measured in the cone coordinate system. The next
 step is to transform the angle $\gamma$ to the position angle measured
 in the plane of the sky. From Equation (1),  neglecting dependence on
 $Z_c$, we can get:
 $$\cot(\psi)={Y_{h}\over Z_{h}}=-\cot(\theta)\sin(\phi)-{Y_{c}\over
 X_{c}}{\cos(\phi)\over\sin(\theta)}. \eqno(10)$$
 Noting that ${Y_{c}\over
 X_{c}}=\cot(90^{o}-\delta)=\tan(\delta)$  we can finally write:
 $$\tan(\delta)=[-\cot(\theta)\sin(\phi)-\cot(\psi)]{\sin(\theta)\over
 \cos(\phi)}. \eqno(11)$$
Now we have all necessary equations to consider the asymmetric cone
model. Using Equations (9) and (11) we can define $\alpha$ as a
function of the position angle ($\psi$). It is important to note
that, to be strictly consistent with the  assumption of constant
velocities of CMEs,  the base of the cone (an ellipse) must be on a
sphere, not on a plane. This needs to introduce a new parameter, a
radius of the sphere, which changes together with an expansion of
CMEs and is very difficult to estimate. Therefore, in our study, we
assume that the base of the cone is a planar ellipse expanding
radially with the same speed everywhere on the ellipse.  This is a
simplification in our model which we believe would not introduce
severe inconsistency in the model.

 Summarizing, we have
seven parameters describing the asymmetric cone model: the expansion
 velocity in 3-D space ($V_{S}$), the angular width of the cone
($\alpha_{\textrm{\tiny{MAX}}}$, measured along the major axis), the
longitude of the cone axis ($\varphi$), the latitude of the cone
axis ($\lambda$), the eccentricity of the cone cross section ($e$),
the orientation of the main axis of the cone cross section ($\beta$)
and finally the localization  of the cone apex (on the surface ($S$)
or at the center of the Sun ($C$)). Using these  parameters we can
determine two more parameters which are typically used to
characterize the cone model: the angle between the central axis of
the cone and the plane of the sky ($\gamma$) and the position angle
of the farthest (fastest-movinng) structure of CME ($PAM$). We
obtain them by the following relationships:
$$\sin\gamma=\sin\vartheta\cos\varphi  \eqno(12)$$
and
$$\tan PAM={\sin\vartheta\cos\varphi \over \cos\vartheta}.   \eqno(13)$$
$PAM$ is measured from solar north in degrees.

\section{Determination of parameters describing HCMEs}
We applied the asymmetric cone model to obtain the  parameters of
frontsided HCMEs recorded by SOHO/LASCO coronagraphs in 2002. We
considered HCMEs originating close to the disk center
($|\varphi|<30^o$) only. These events are clear HCMEs (in contrast
to limb events  appearing as halos due to  deflections of
pre-existings coronal structures) and they   are sufficiently bright
to obtain height-time plots around the entire occulting disk. Two
main reasons for looking at these events more closely are: (1) they
are responsible for severe geomagnetic storms and (2) not all such
symmetric events were considered by Michalek \emph{et al.} (2003).
A two-step procedure was carried out to obtain the parameters
characterizing HCMEs. First, using the height-time plots the
projected speeds at different position angles were determined. We
made measurements, for the considered events, every $15^{o}$ around
the entire occulting disk. This allowed us to obtain $24$ points for
each event which are required for the fitting procedure. Second,
using numerical simulation through minimizing the root mean square
error the cone model parameters were obtained.  The simulation
procedure was also performed in two steps. First, we executed
simulations without any constraints on the cone parameters. These
simulations performed with small accuracy produced only approximate
values for the cone parameters.  Next, more precise simulations
(with constraints on the cone parameters) with better accuracy were
made to obtain the final  best-fit cone model parameters. To compare
the asymmetric (new) and symmetric (standard) cone model we
performed both types of simulations. The results of our studies are
presented in the next section.

\section{Results}
 During 2002 we selected 15 frontsided HCMEs originating close
to the disk center ($|\varphi|\leq 30^o$). For all of them,   the
cone parameters using the asymmetric (ACM) and symmetric cone models
(SCM) were determined. The results for the ACM and SCM are presented
in Tables~I~and~II, respectively. The
 first four columns are from the SOHO/LASCO catalog and give the date, time of the first appearance
 in the coronagraph field of view,
 projected speed and position angle of the farthest (fastest-moving)
  part of the HCME. In column (5) the source locations of the
 associated H-alpha flares are presented. Parameters $\alpha$, source
locations and $V_{S}$ estimated from the  cone model are shown in
columns (6), (7), and (8), respectively. In column (9) the r.m.s
errors (in km~s$^{-1}$) for the best fits are given. The parameters
$\gamma$ and $PAM$ are shown in columns (10) and (11), respectively.
These data are presented in  both tables. In  Table~I, we have three
additional columns displaying the new parameters used only for the
asymmetric cone model. In  columns (12) and (13),  the eccentricity
($e$) and orientation of the  cone base ($\beta$) are presented. In
the last column (14) the location of the cone apex is described
($S$: surface of the Sun, $C$:  center of the Sun). Figures~4-17 are
polar plots showing the  fitted projected speeds (solid lines) and
measured projected speeds (cross symbols) in the velocity space.
Left and right panels are for    the ACM and the SCM models,
respectively. From the polar plots and the r.m.s errors presented in
the tables it is evident  that  the ACM fits much better the
measured projected speeds in comparison with the SCM. For all
considered events the r.m.s errors are much smaller for the ACM than
for the SCM. The average r.m.s errors for all events are equal to
60~km~s$^{-1}$ and 84~km~s$^{-1}$ for the ACM and SCM, respectively.
If CMEs could be considered as the symmetric cones then their
projected speed should also follow the  elliptic shapes in the plane
of the sky. Checking Figures~4-17, we see that the measured
projected speeds for all considered events show more complicated
structures which can not be fitted by the SCM. Only the ACM  can
adjust shapes indicated by the measured projected speed with good
accuracy. From the inspection of the tables, we see that the cone
parameters obtained from  both models are slightly different. On
average, the space velocities are about 50~km~s$^{-1}$ higher and
the cone widths are about $4^o$ lower for the SCM in comparison with
the ACM. If we consider the ACM only, it is evident that more events
(11 out of 15) have better fits when the cone apex is placed in the
center of the Sun. This means that CMEs are likely to originate from
the finite area but not from   the point source. It is important to
note, that all events with velocities above 1200~km~s$^{-1}$ have
better fits when the cone apex is placed at the center of the Sun.
It seems obvious that powerful events need more energy input. The
eccentricity ($e$) and orientation of the cone base ($\beta$), for
the considered events, vary in a random way between $0.8-0.4$ and
$0^{o}-180^{o}$, respectively. Using our model (working in the
velocity space) we can also reconstruct the space shapes (in the
plane of sky) of CMEs. Figure 17 shows an example of LASCO CME (halo
CME from 2002/03/15) with  dark dots representing the projected
radial distances derived from the ACM. To obtain these dots we
assumed that the onset of this CME was at 22:09UT (X-ray flare
onset) on 15 March 2002. We can see that the model reproduces the
space shape of the CME very well.
\section{Summary}
In this paper we have presented the new asymmetric cone model  to
determine the geometrical  parameters characterizing  HCMEs. We
applied this model to all  frontsided HCMEs originating close to the
disk center and listed in the SOHO/LASCO catalog in 2002. We
estimated, with very good accuracy (the average r.m.s error is equal
to 60~km~s$^{-1}$), the cone parameters for $15$ events. It was
shown (from the polar plots and the r.m.s errors) that the ACM can
reproduce, around the entire occulting disk, the measured projected
speed much better than the SCM. We have to note that sometimes the
individual projected speeds may significantly differ from the
fitting model even for the ACM. In these cases errors are probably
caused by inaccurate measurements. Unfortunately, most HCMEs are not
sufficiently bright around the entire occulting disk to generate the
height-time plots with the same good accuracy for  all considered
position angles. The biggest errors may arise when the faintest part
of a given event is considered or when measurements are disturbed by
 another event appearing in the LASCO field of view. We also
confirmed that most of the events have better fits when the cone
apex is located at the center of the Sun.  We used this model for
the frontsided full HCMEs originating close to the disk center but
it can also be applied to limb and even partial HCMEs.  When
considering limb or partial HCMEs the accuracy will be slightly
worse. The determined  parameters for HCMEs are similar to that
derived by the other cone models, but for some events differences
could amount to even 20\%.  We have to remember that this approach
has two shortcomings: (1) CMEs may be
 accelerating, moving with constant speed or decelerating at the
 beginning phase of propagation. This means that the constant velocity assumption
may be invalid.(2) CMEs may expand in addition to
 radial motion. Then the measured sky-plane speed is a sum of the
 expansion speed and the projected radial speed. This would also
 imply that  CMEs may not be a rigid cone as we had assumed. Unfortunately,  having observation from only one
 spacecraft there is no possibility to overcome these assumptions.
 There are two attempts  to get the real space parameters of CMEs.
 Some scientist
   (e.g. Xie \emph{et al.}, 2004) try to get the real parameters of CMEs by considering the elliptical shapes of
   observed CMEs.  In our approach we consider velocities of
   CMEs as a function of the position angle. We take into account
   the projected velocities (many points) from an entire halo CME, not only from a few chosen  points.
   We do not introduce any assumption about the apparent (geometrical) shapes of CMEs.
   Therefore, we expect that our determination of parameters may have  better accuracy.
The determined real velocities of halo CMEs could be potentially
useful for space weather forecast. But we have to note that
although it is true that faster interplanetary CMEs are more
geoeffective, the strength and topology of magnetic field is still
crucial to define geoeffectiveness of CMEs (e.g., Bothmer, 2003).

\begin{acknowledgements}
 This paper is based on the author's work at Center
 for Solar and Space Weather,
Catholic University of America in Washington. In this paper we used data from SOHO/LASCO CME catalog. This CME
catalog is generated and maintained by the Center for Solar Physics and Space Weather,
 The Catholic University of America in cooperation with the Naval
Research Laboratory and NASA. SOHO is a project of international
cooperation between ESA and NASA. The author thanks N. Gopalswamy,
S. Yashiro and H. Xie for helpful discussions during work on the
manuscript. This wark was  supported by {\it Komitet Bada\'{n}
Naukowych} through the grant PB~0357/P04/2003/25 and NASA program
LWS.
\end{acknowledgements}

\newpage

%\baselineskip=0.5

\begin{figure*}
\vspace{7.0cm} \includegraphics{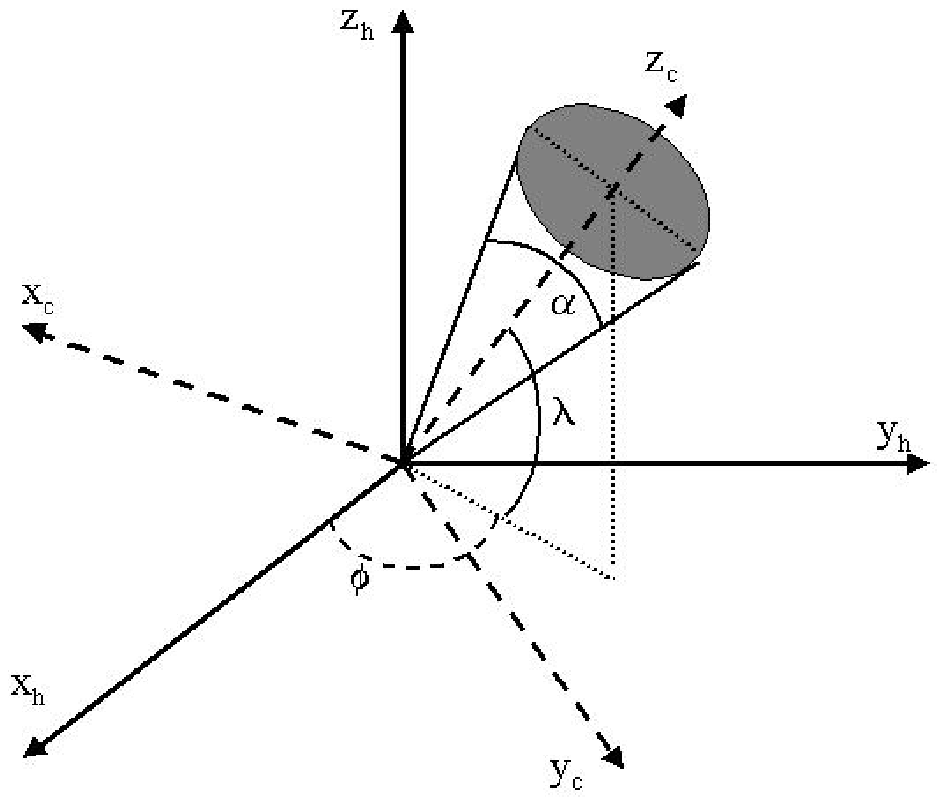} \vspace{0mm}\caption{The cone model
topology and relationship between the heliocentric coordinate system
and the cone coordinate system when the cone apex is placed at the
center of the Sun.}
\end{figure*}

\begin{figure*}
\vspace{7.0cm} \includegraphics{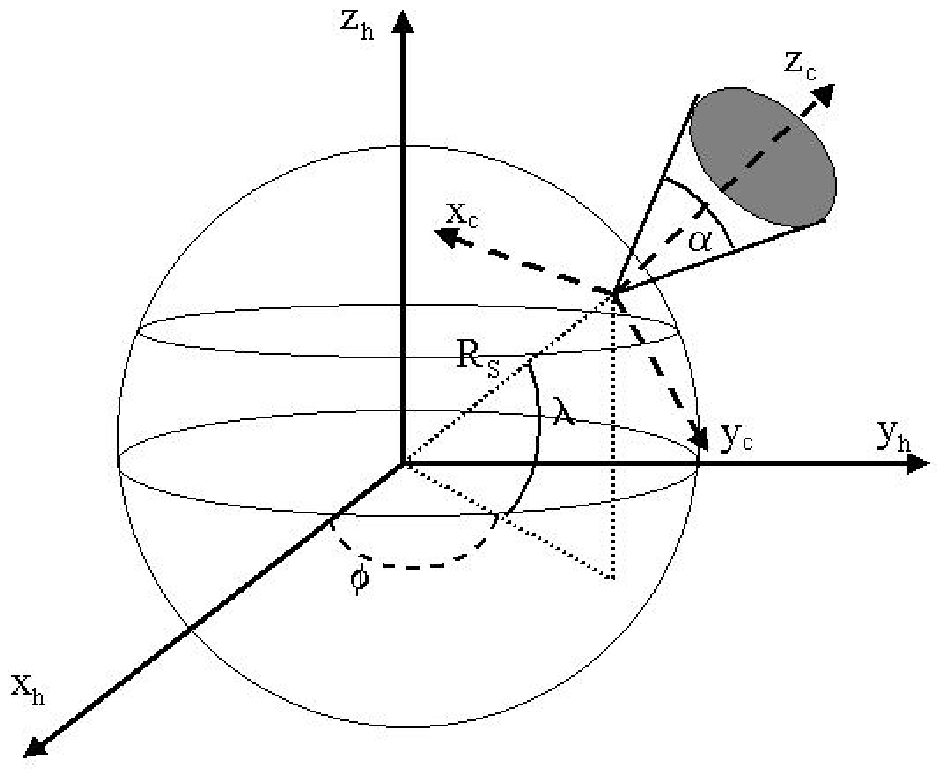} \vspace{0mm}\caption{The cone model
topology and relationship between the heliocentric coordinate system
and the cone coordinate system when the cone apex is placed on the
surface of the Sun.}
\end{figure*}

\begin{figure*}

\vspace{7.0cm} \includegraphics{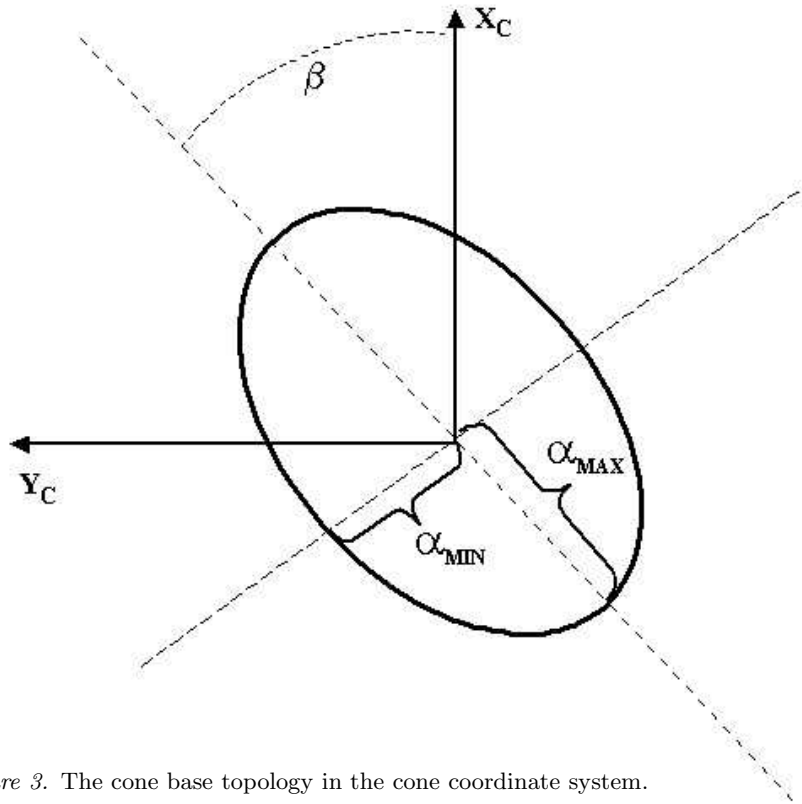} \vspace{0mm}\caption{The cone  base
topology in the cone coordinate system.}

\end{figure*}

\begin{figure*}
\vspace{7.0cm} \includegraphics{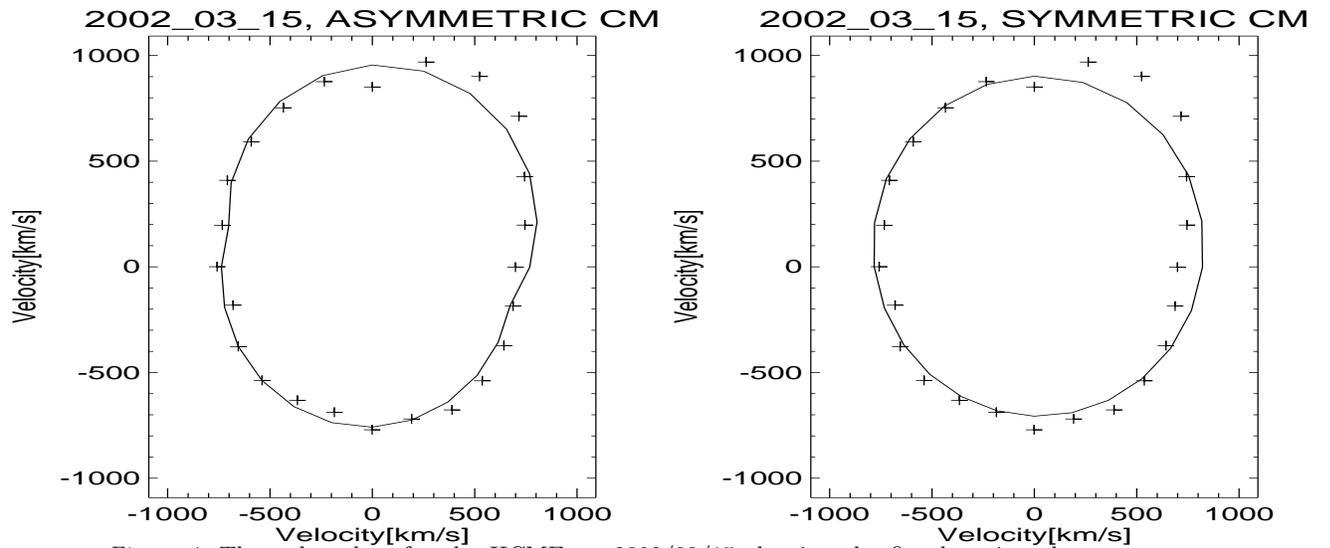} \vspace{0mm}\caption{The polar
plots for the HCME on 2002/03/15 showing the fitted projected speeds
(solid lines) and measured projected speeds (cross symbols) in the
velocity space.  For comparison the best fits for the asymmetric
cone model (left panel) and the symmetric cone model (right panel)
are presented. }
\end{figure*}

\begin{figure*}
\vspace{7.0cm} \includegraphics{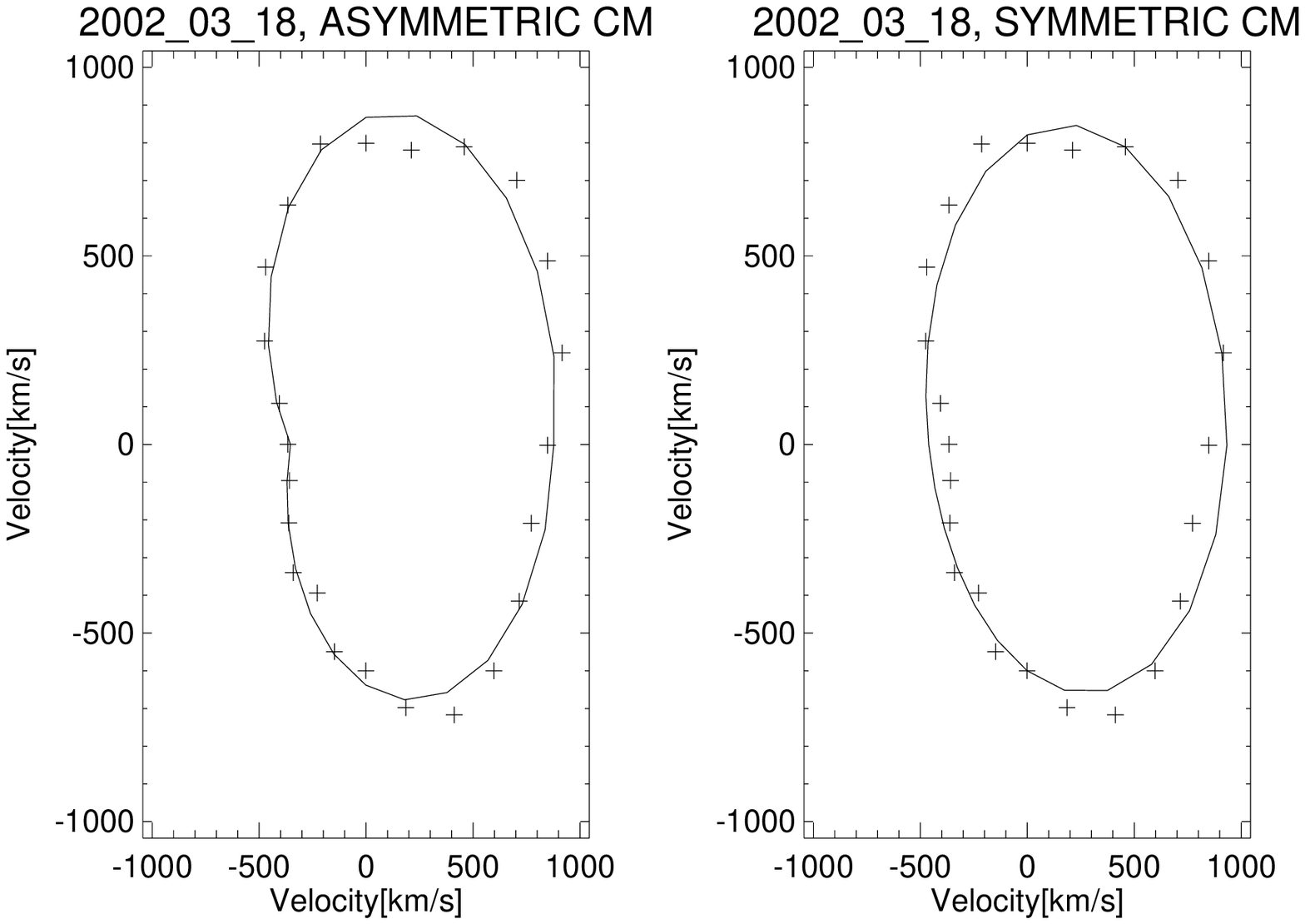} \vspace{0mm}\caption{The polar
plots for the HCME on 2002/03/18. The format is the same as in
Figure 4.}
\end{figure*}

\begin{figure*}
\vspace{7.0cm} \includegraphics{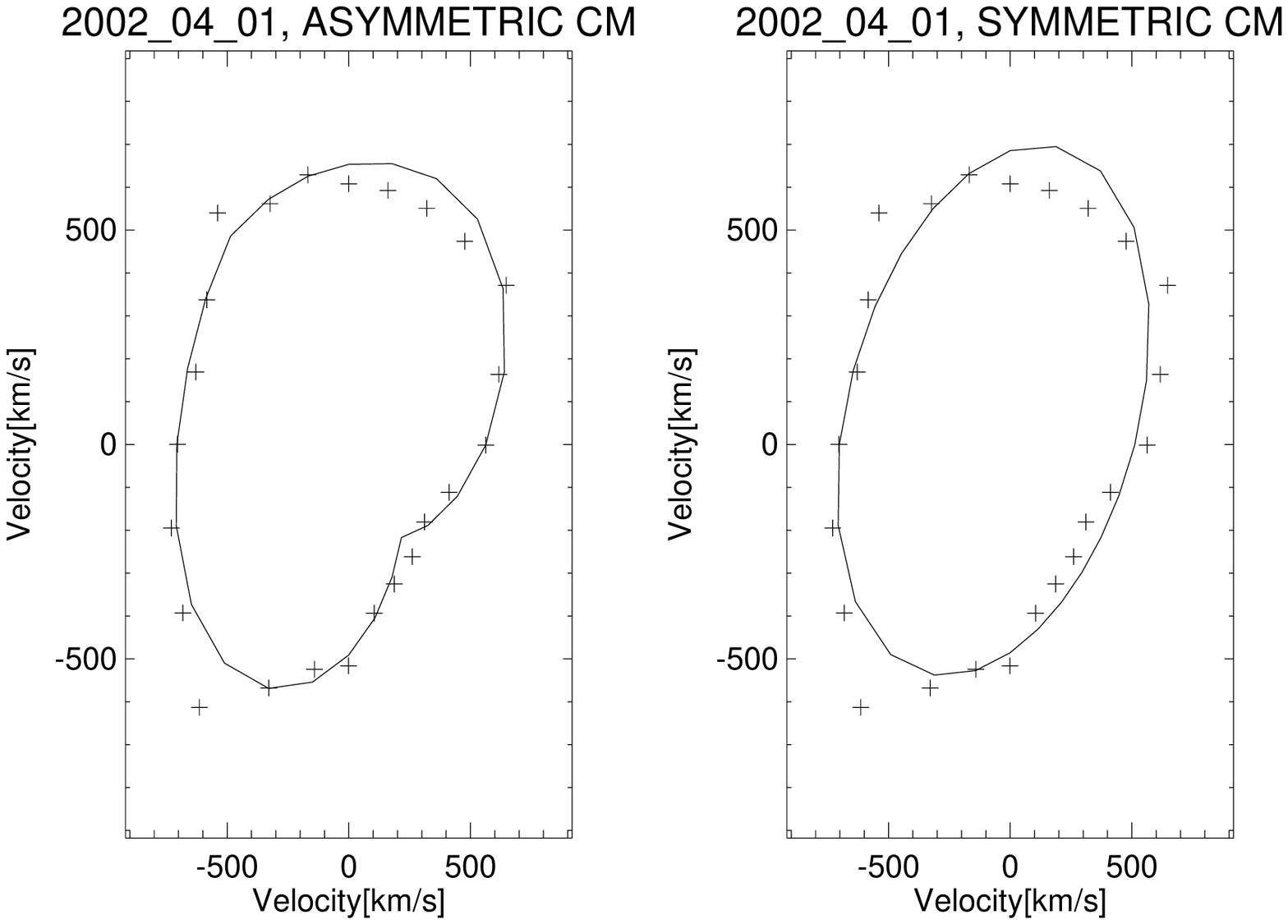} \vspace{0mm}\caption{The polar
plots for the HCME on 2002/04/01. The format is the same as in
Figure 4. }
\end{figure*}

\begin{figure*}
\vspace{7.0cm} \includegraphics{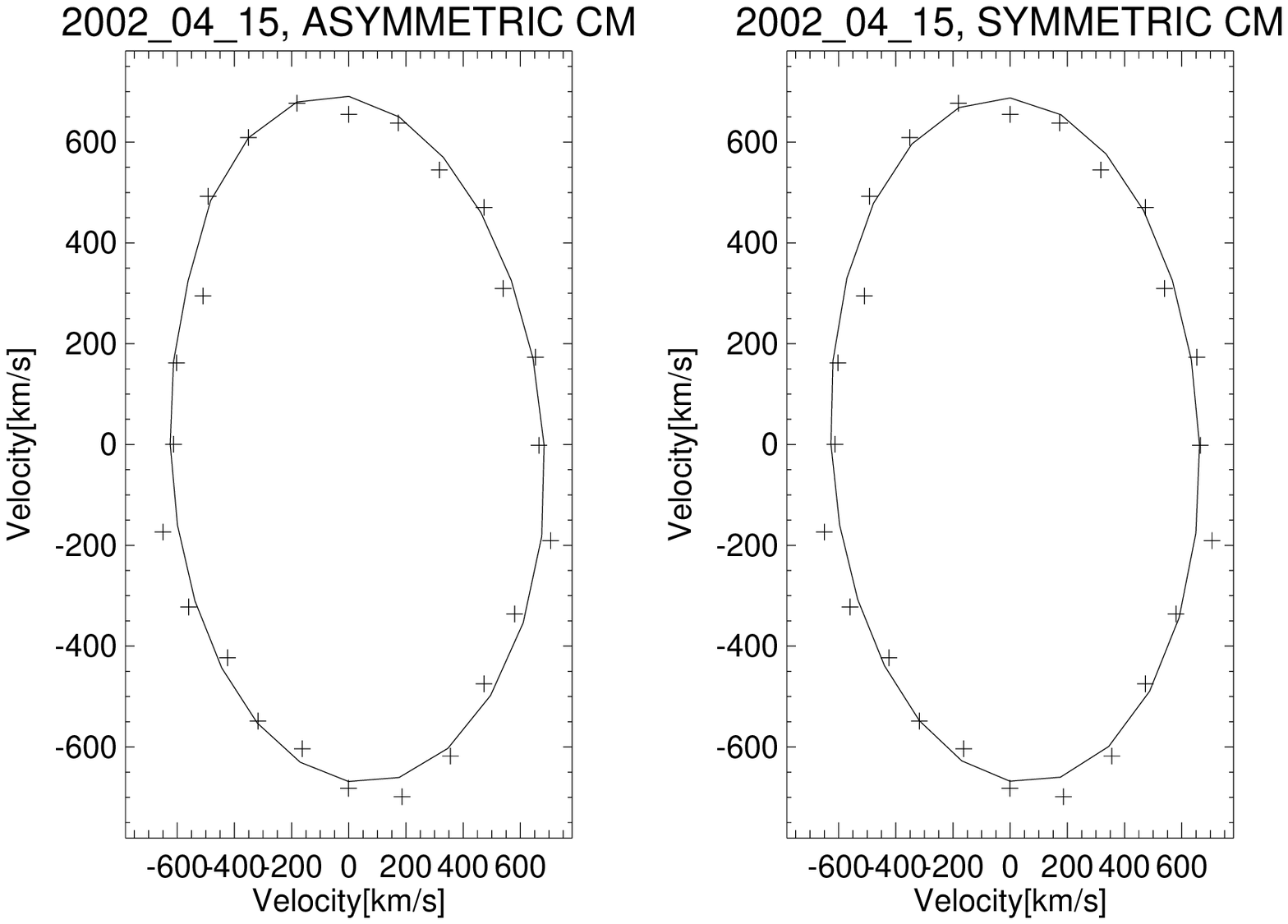} \vspace{0mm}\caption{The polar
plots for the HCME on 2002/04/15. The format is the same as in
Figure 4.  }
\end{figure*}

\begin{figure*}
\vspace{7.0cm} \includegraphics{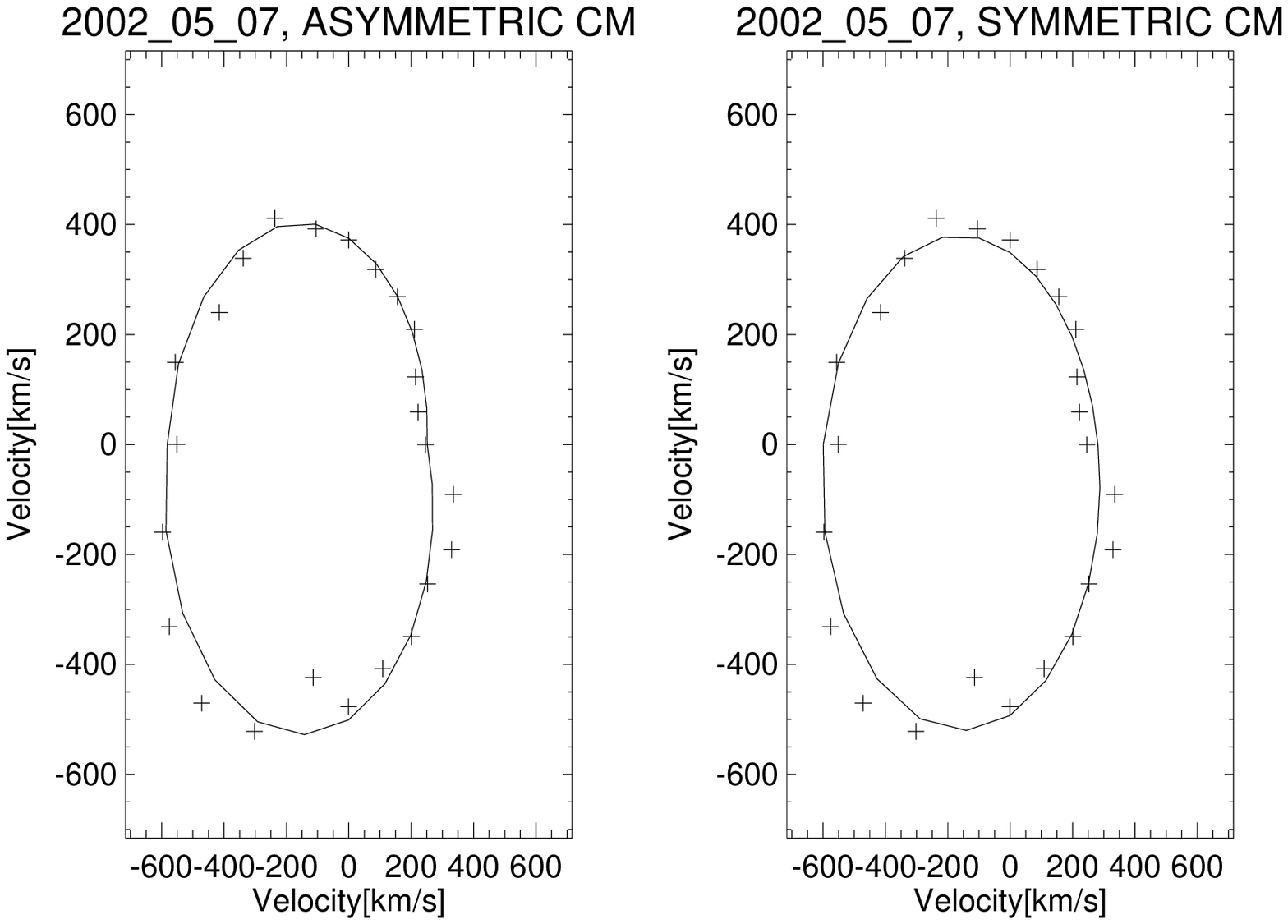} \vspace{0mm}\caption{The polar
plots for the HCME on 2002/05/07. The format is the same as in
Figure 4. }
\end{figure*}

\begin{figure*}
\vspace{7.0cm} \includegraphics{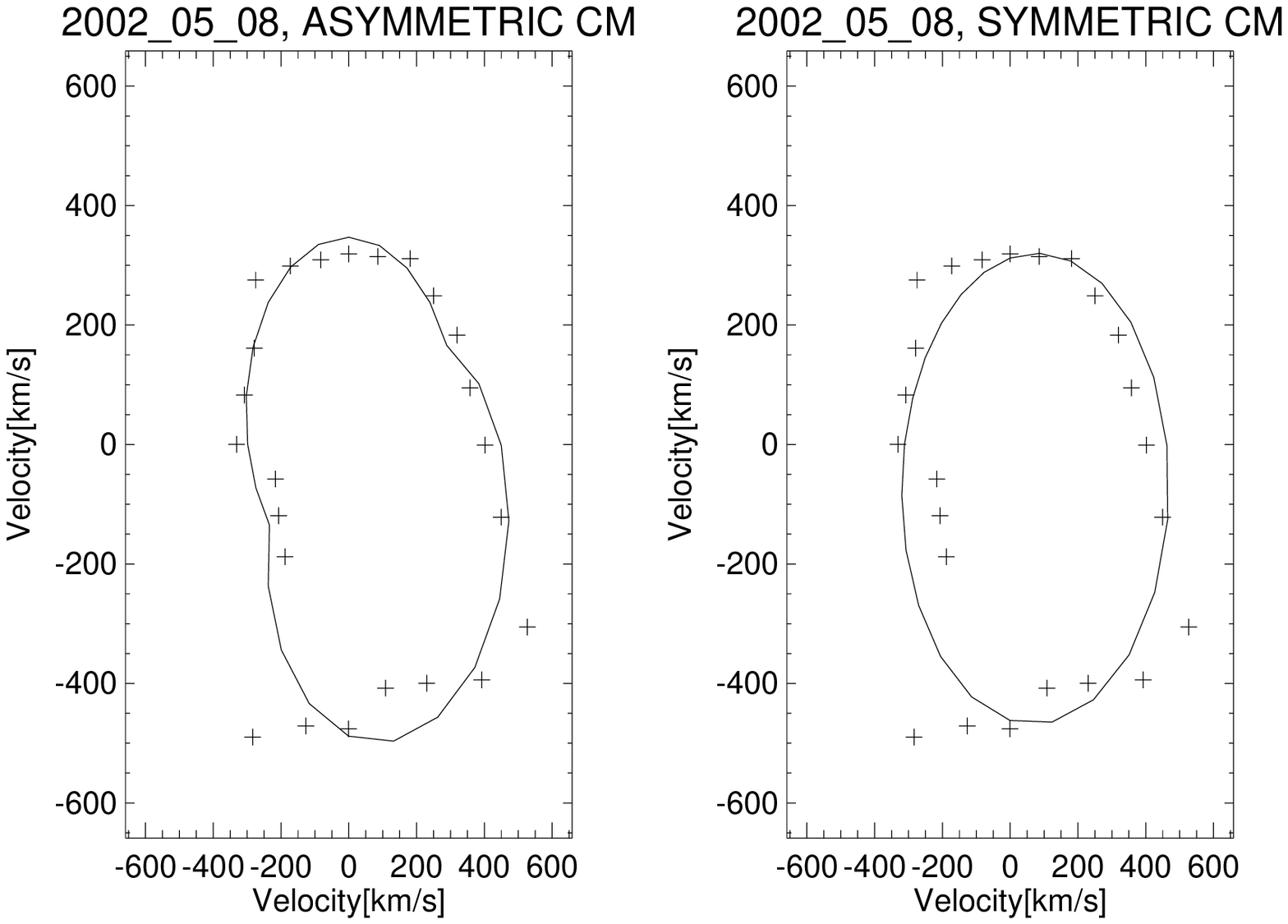} \vspace{0mm}\caption{The polar
plots for the HCME on 2002/05/08. The format is the same as in
Figure 4. }
\end{figure*}

\begin{figure*}
\vspace{7.0cm} \includegraphics{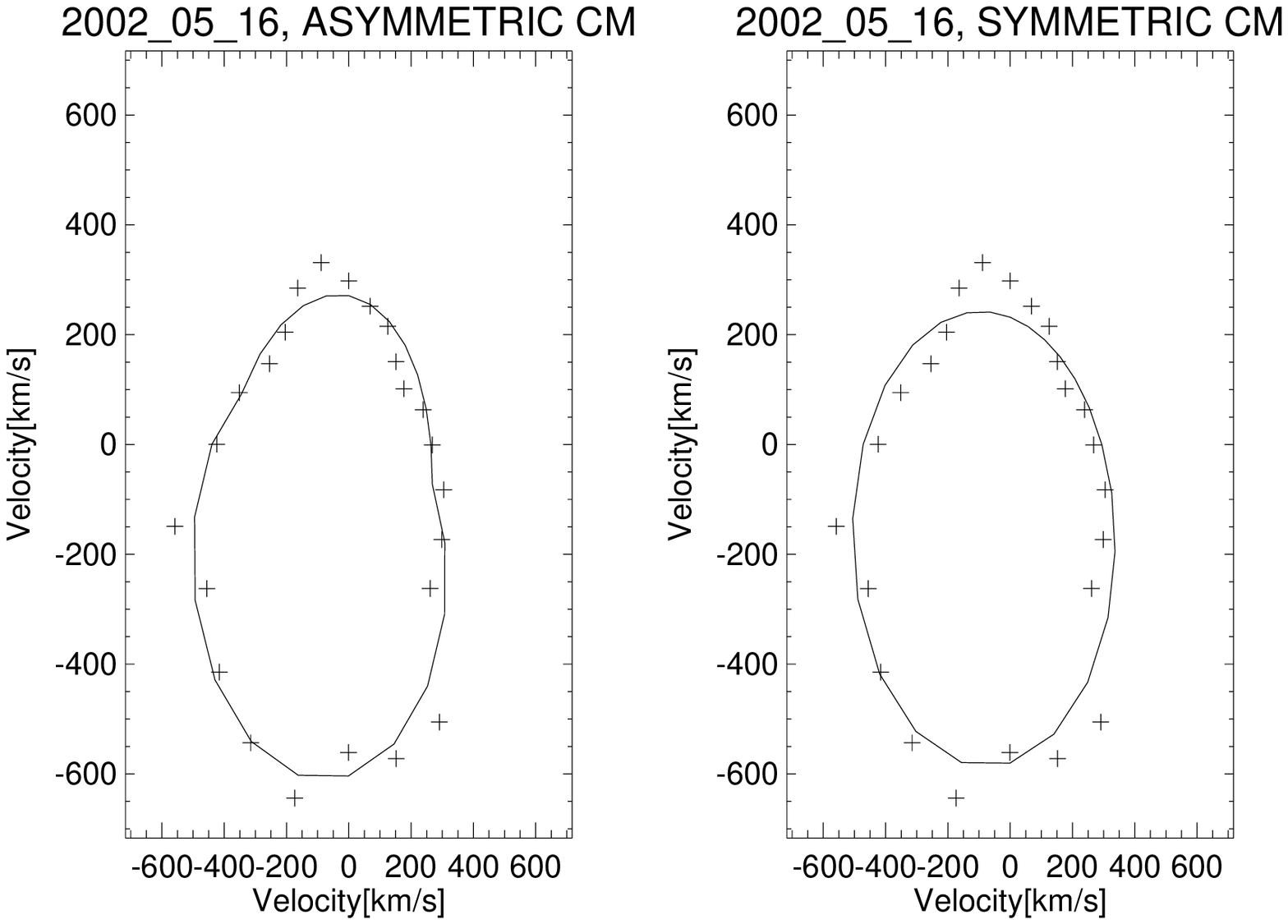} \vspace{0mm}\caption{The polar
plots for the HCME on 2002/05/16. The format is the same as in
Figure 4.  }
\end{figure*}

\begin{figure*}
\vspace{7.0cm} \includegraphics{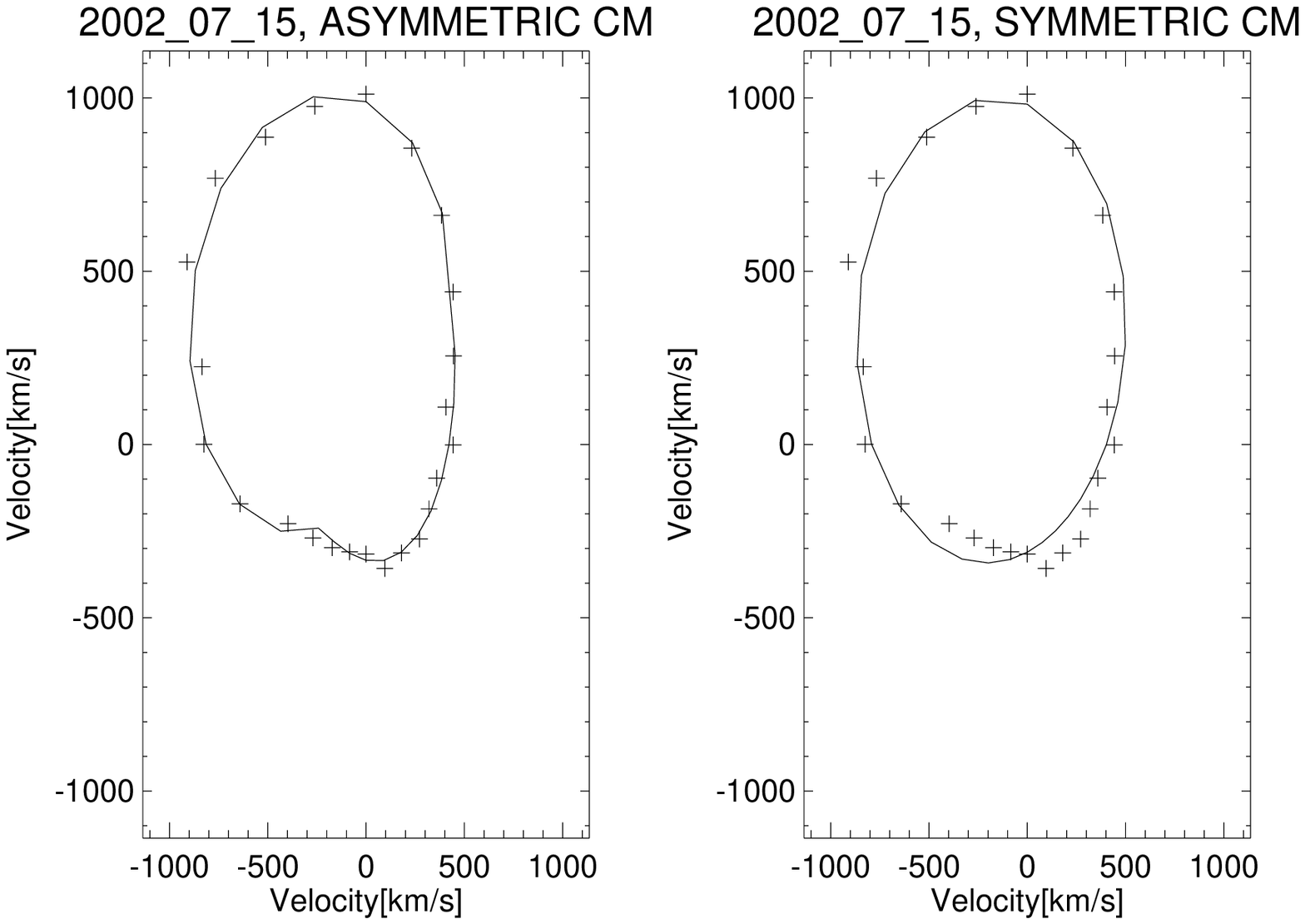} \vspace{0mm}\caption{The polar
plots for the HCME on 2002/07/15. The format is the same as in
Figure 4. }
\end{figure*}

\begin{figure*}
\vspace{7.0cm} \includegraphics{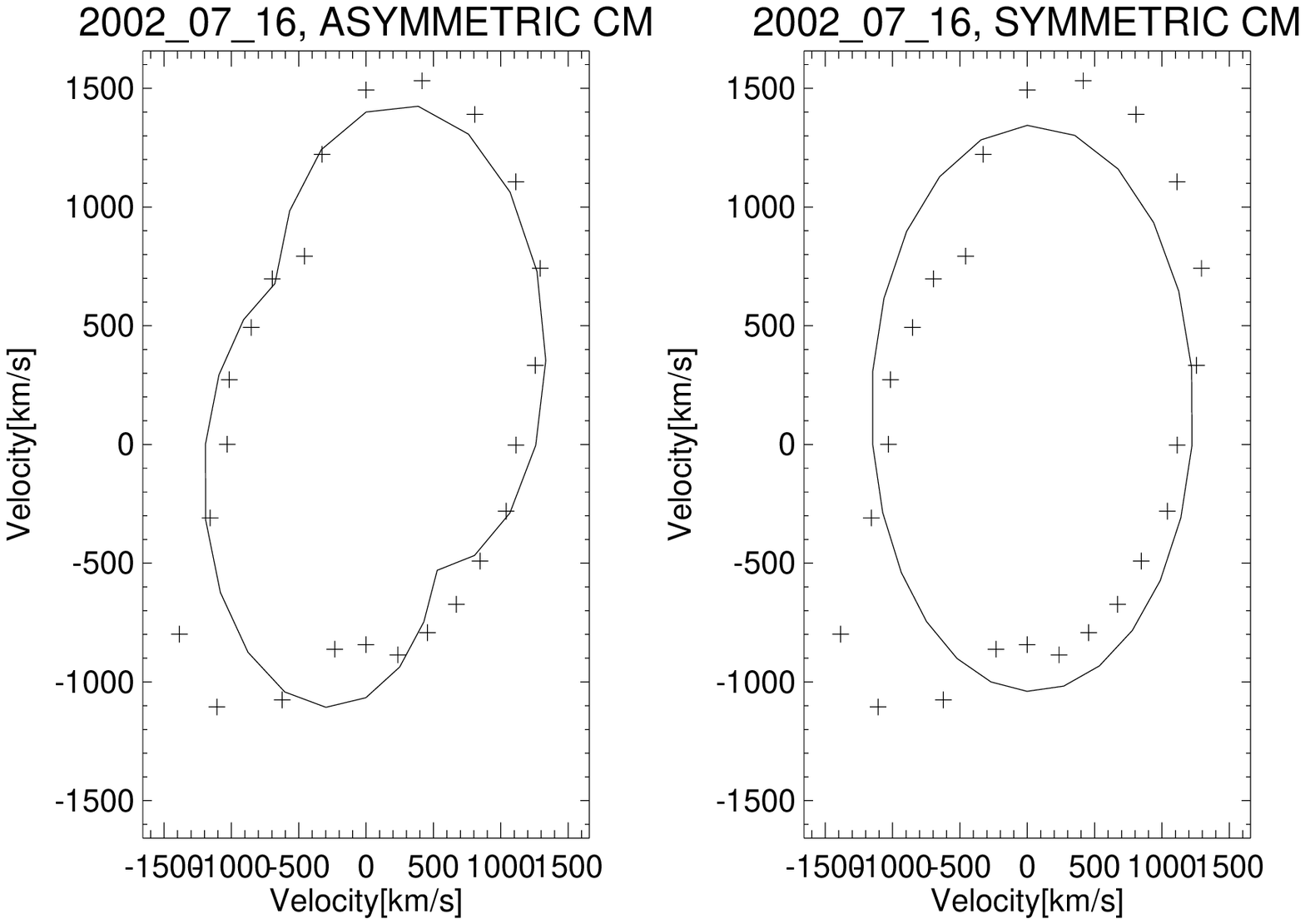} \vspace{0mm}\caption{The polar
plots for the HCME on 2002/07/16. The format is the same as in
Figure 4. }
\end{figure*}

\begin{figure*}
\vspace{7.0cm} \includegraphics{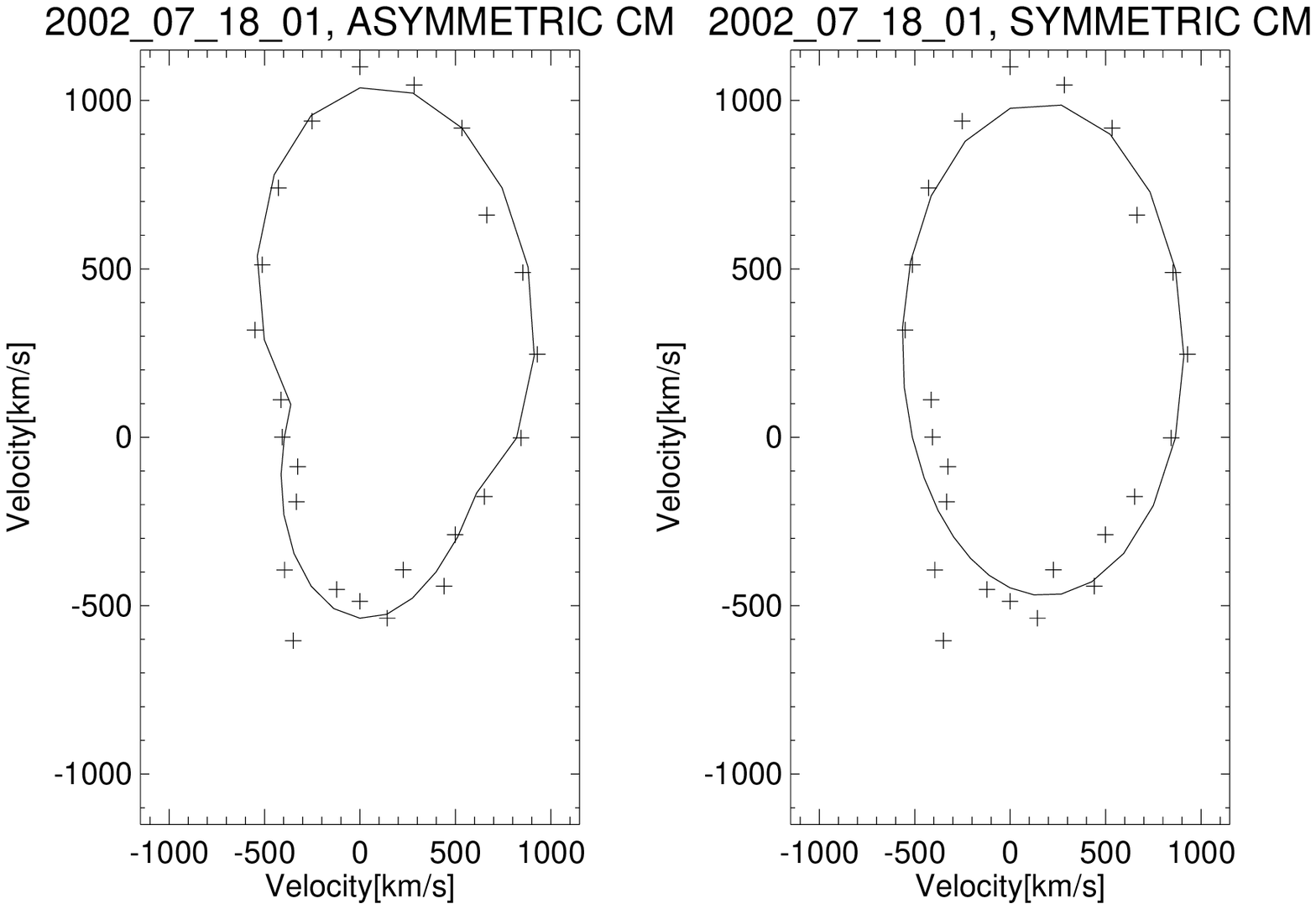} \vspace{0mm}\caption{The polar
plots for the HCME on 2002/07/18. The format is the same as in
Figure 4. }
\end{figure*}

\begin{figure*}
\vspace{7.0cm} \includegraphics{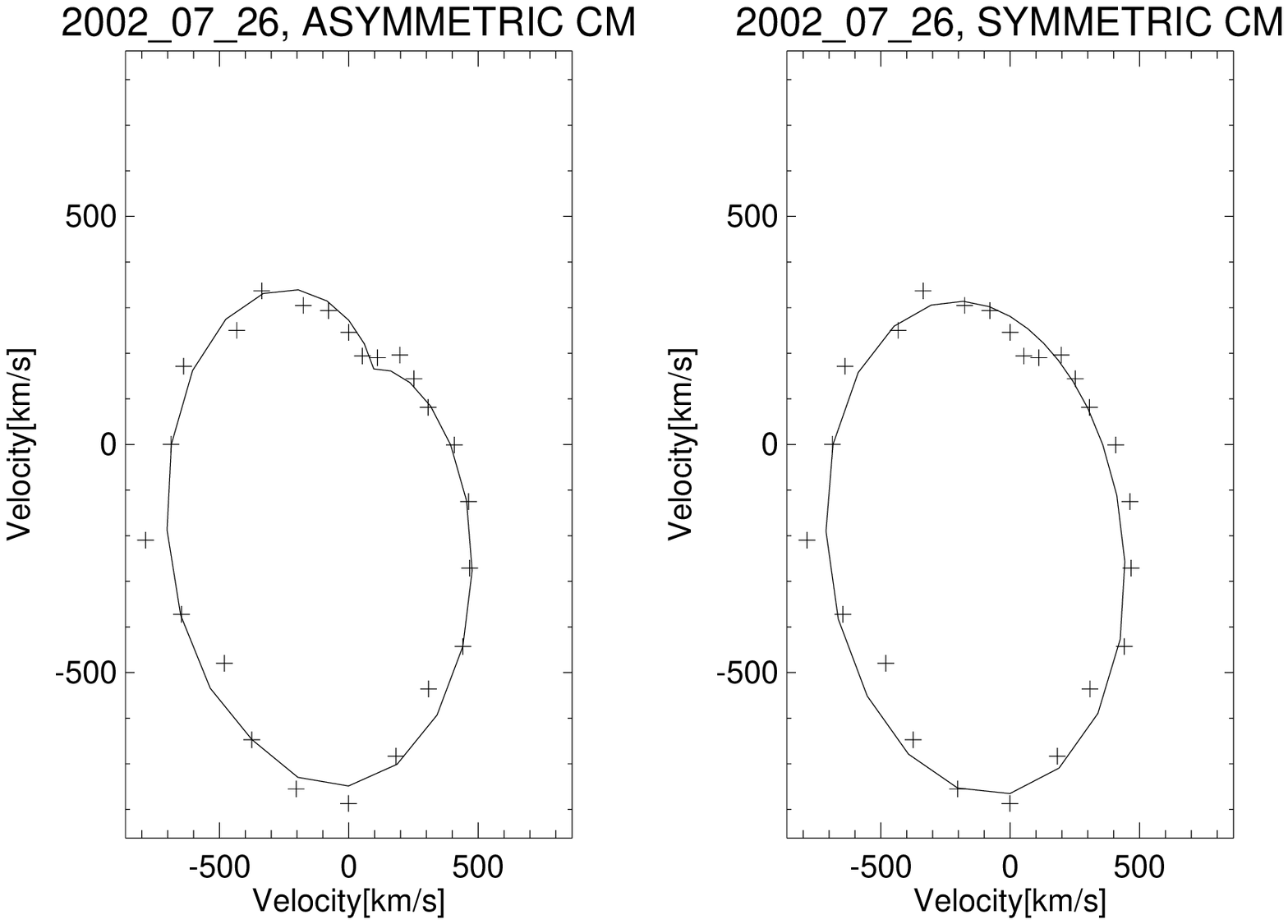} \vspace{0mm}\caption{The polar
plots for the HCME on 2002/07/26. The format is the same as in
Figure 4.  }
\end{figure*}

\begin{figure*}
\vspace{7.0cm} \includegraphics{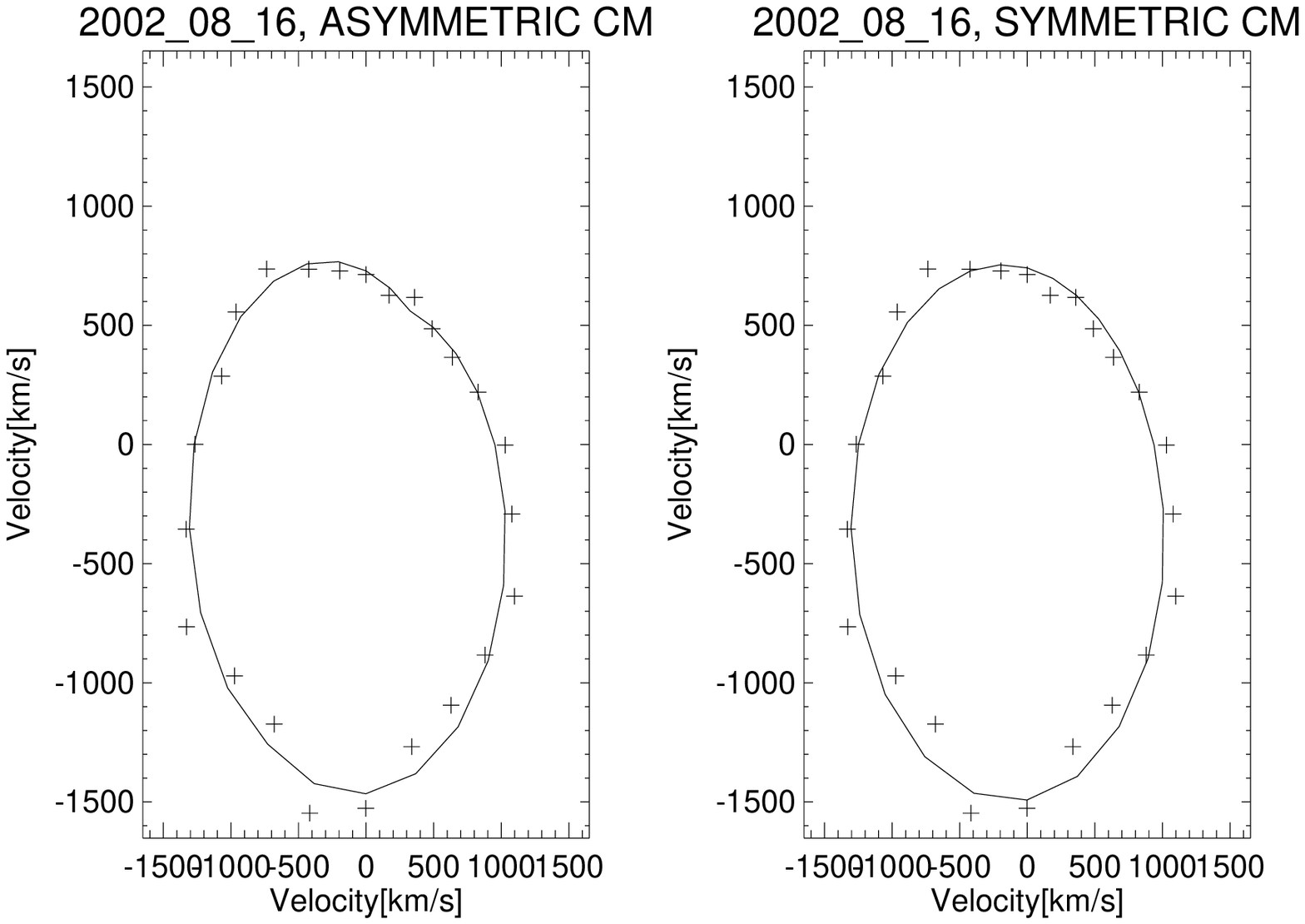} \vspace{0mm}\caption{The polar
plots for the HCME on 2002/08/16. The format is the same as in
Figure 4.  }
\end{figure*}

\begin{figure*}
\vspace{7.0cm} \includegraphics{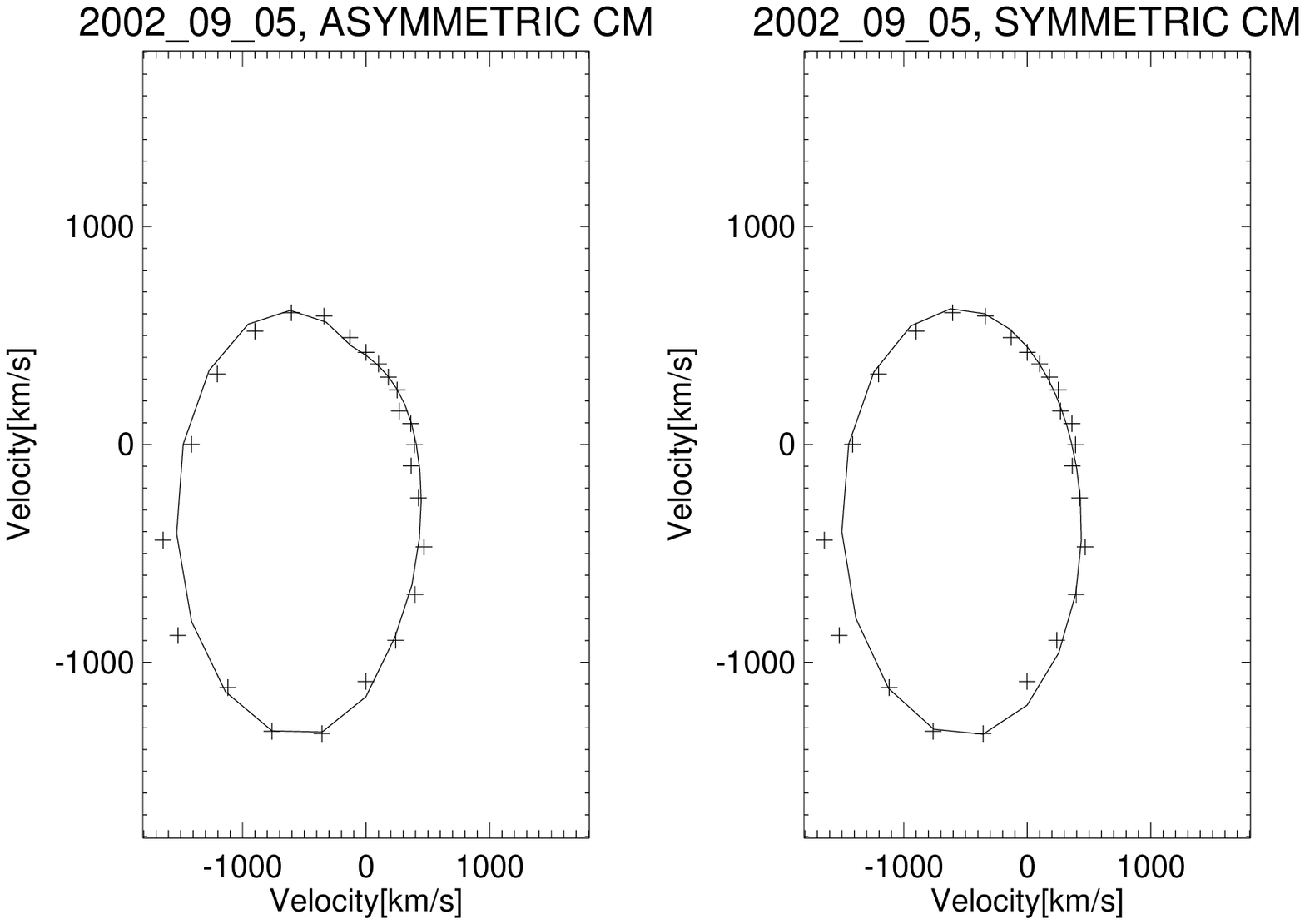} \vspace{0mm}\caption{The polar
plots for the HCME on 2002/09/05. The format is the same as in
Figure 4.  }
\end{figure*}

%\begin{figure*}
%\vspace{17.0cm} \special{psfile=2002_11_09praca.ps hscale=100
%vscale=58 hoffset=-100 voffset=-220}
% \vspace{0mm}\caption{The polar
%plots for the HCME on 2002/11/09 showing the fitted projected speeds
%(solid lines) and measured projected speeds (cross symbols) as
%function of heliolatitude. For comparison the best fits for the
%asymmetric cone model (left panel) and the symmetric cone model
%(right panel) are presented. }
%\end{figure*}

%\begin{figure*}
%\vspace{17.0cm} \special{psfile=2002_12_19_02praca.ps hscale=100
%vscale=58 hoffset=-100 voffset=-220} \vspace{0mm}\caption{The polar
%plots for the HCME on 2002/12/19 showing the fitted projected speeds
%(solid lines) and measured projected speeds (cross symbols) as
%function of heliolatitude. For comparison the best fits for the
%asymmetric cone model (left panel) and the symmetric cone model
%(right panel) are presented. }
%\end{figure*}
%\end{document}

\begin{figure*} \vspace{9.0cm} \includegraphics{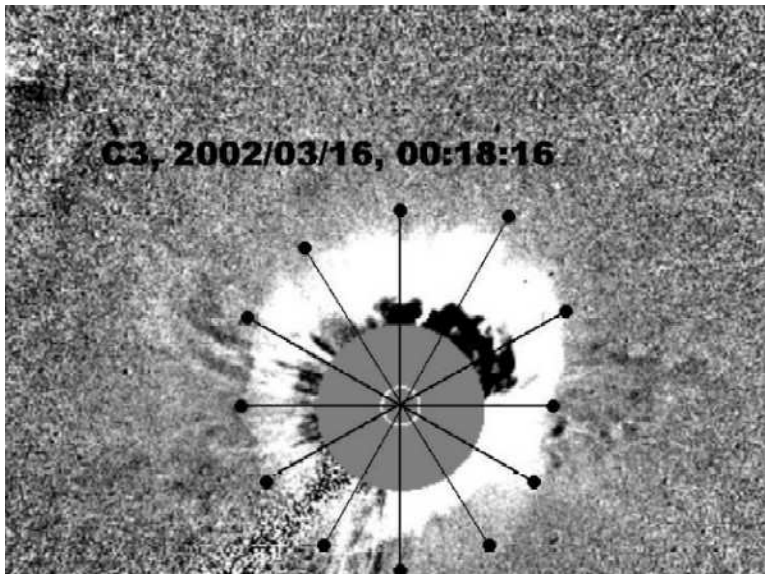}
\vspace{0mm}\caption{The halo CME recorded by  LASCO/C3 coronagraph
 on 16 March 2002 at 00:18 UT (In LASCO catalog it is halo CME from
2002/03/15). Dark dots represent radial  distances derived for this
event from ACM. }
\end{figure*}

%
%\end{document}

%\end{document}

\begin{table*}
%{\scriptsize
 %  The Table~1 shows the list of frontsided halo CMEs recorded by SOHO/LASCO (in 2002)
%coronagraphs originating close to the solar center $(|\varphi|<30^o)$. The
 %first four columns are from the SOHO/LASCO catalog and give date, time of first appearance
 %in the coronagraph field of view,
 %projected speed and position angle of the fastest part of the HCME.
%In column (5) the source locations of the associated H-flares are presented. Parameters $\alpha$, source
%locations and $V_{S}$ estimated from the assymmetric cone are shown in columns (6), (7), and (8), respectively.
%In column (9) the r.m.s errors for the best fit are displayed. The parameters $\gamma$ and $PAM$ are given in
%columns (10) and (11), respectively. In the columns (12) and (13) parameters describing eccentricity ($e$) and
%orientation of the base ($EO$)of the cone are displayed. In the last column (14) the localization of the cone
%apex is described ($S$-surface of the Sun, $C$- center of the Sun).}\\
%{  { List of halo CMEs with parameters received form the ACM}}
%{\label List of halo CMEs.}\\
{ \normalsize \textbf{Table~I}\\
List of halo CMEs with parameters derived from the ACM.}
{\tiny

\begin{tabular}{c c c c c c c c c c c c c c }

 \hline \hline

Date & Time & Speed & PA & Flare  & $\alpha_{\tiny MAX}$ & Location & V & Error   &  $\gamma$ &  PAM & e &  $\beta$  &   Cone Apex  \\

\hline
    &    &   km~s$^{-1}$  &  Deg &     &  Deg  &      &  km~s$^{-1}$ & km~s$^{-1}$  &   Deg &  Deg &  &  Deg  &   \\

\hline

2002/03/15 & 23:06:06 & 957 &  309 & S08W03 &  95 &  N07W01  &   1178  &   47  &  81 & 347  &   0.6  &   75   &    S   \\
2002/03/18 & 02:54:06 & 989  & 311 & S04W24 &  114 &  N11W22  &   947  &   42  &  64  & 296  &   0.6 &    90  &     C  \\
2002/04/01 & 13:25:05 & 474  &  350 &  N11E21 & 167 &   N35E30 &  747  &    46  &  46  &  35  &    0.6&    45  &    C  \\
2002/04/15 & 03:50:05 & 720  &  198 &  S18W01 & 168 &  N14W21 &  715  &    27  &  66  & 304  &   0.4&   60   &      C \\
2002/05/07 &04:06:05  & 720  & 112 & S22E14  & 44  &   S03E08   &  1258  &   39   & 81   &110&      0.4&    90  &    C   \\
2002/05/08 &13:50:05  & 614  & 229  &S12W07  & 77 &    S08W08  &  657  &   55  &  78  &224  &    0.7 &   120  &      S \\
2002/05/16 &00:50:05  & 600   &158  &S22E14  & 60  &    S12E06 &   915  &   38  &  77  &153 &     0.6 &    75   &    C  \\
2002/07/15 &20:30:05  &1151  & 35   &N19W01  & 67  &    N18E10 &   1249 &   28  &  69  &  28 &    0.6 &   135  &     C  \\
2002/07/16 &16:02:58  &1636  &325   &N23W07   & 70 &   N05W01  &   2268 &   150 &  85  & 348 &     0.8&   45   &    C  \\
2002/07/18 &08:06:08  &1099   &354   &N19W30  & 104 &    N20W13 &  1110  &  66  &  66  & 328 &    0.7 &    75  &    S \\
2002/07/26  &22:06:10  & 818  &172   &S19E26  & 80  &    S21E13 &   846  &  36  &  65  & 149 &     0.6&  150   &    S \\
2002/08/16 & 12:30:05  &1585  &121   &S14E20   & 72 &     S14E06 &  1895 &  68  &  74  & 157 &    0.4 &  150   &   C \\
2002/09/05  &16:54:06  &1748  &114   &N09E28  & 41  &    S08E12  &  2758 &  52  &  75  & 123 &    0.4 &   15   &    C \\
2002/11/09 & 13:31:45  &1838  &233   &S12W29  & 52  &    S08W09  &  2658 & 100  &  78   &228  &   0.6 &  105   &   C \\
2002/12/19 &22:06:05 &1092    &300  &N15W09   & 67  &    N06W01  &  2504 & 110  &   84 &  352 &   0.7 &  45   &   C  \\

 \hline
\end{tabular}
}
%\end{table*}
%\end{document}

%\begin{table*}

%{\scriptsize The Table~2 shows the list of frontsided halo CMEs recorded by SOHO/LASCO (in 2002) coronagraphs
%originating close to the solar center $(|\varphi|<30^o)$. The
% first four columns are from the SOHO/LASCO catalog and give date, time of first appearance
% in the coronagraph field of view,
% projected speed and position angle of the fastest part of the HCME.
%In column (5) the source locations of the associated H-flares are presented. Parameters $\alpha$, source
%locations and $V_{S}$ estimated from the symmetric cone are shown in columns (6), (7), and (8), respectively. In
%column (9) the r.m.s errors for the best fit are displayed. The parameters $\gamma$ and $PAM$ are given in
%columns (10) and (11), respectively.}\\
{\normalsize \textbf{Table~II}\\
List of halo CMEs with parameters derived from
the SCM.}
%{\label List of halo CMEs.}\\

{\tiny

\begin{tabular}{c c c c c c c c c c c }

 \hline \hline

Date& Time & Speed & PA & Flare  & $\alpha$ & Location & V & Error   &  $\gamma$ &  PAM    \\

\hline
    &    &   km~s$^{-1}$  &  Deg &     &  Deg  &      &  km~s$^{-1}$  & km~s$^{-1}$ &   Deg &  Deg   \\

\hline
2002/03/15 & 23:06:06 & 957 &  309 & S08W03 & 72  &   N05W01 &    1375  &  62 &  85 &   348 \\
2002/03/18 & 02:54:06 & 989  & 311 & S04W24 &121  &   N09W23  &    969  &  57 &  65  &  292\\
2002/04/01 & 13:25:05 & 474  &  350 &  N11E21 &156 &  N31E32   &   738   & 66 &  46  &   41\\
2002/04/15 & 03:50:05 & 720  &  198 &  S18W01 & 169 &  N10W21   &   692 &  30  &  67 &   296\\
2002/05/07 &04:06:05  & 720  & 112 & S22E14  & 48  &   S04E09  &  1110  &   38 & 80  &  113\\
2002/05/08 &13:50:05  & 614  & 229  &S12W07  & 58  &   S06W06  &   818   &  69 & 81   & 225\\
2002/05/16 &00:50:05  & 600   &158  &S22E14  & 54   &  S12E06 &   934  &   49 & 77  &  153\\
2002/07/15 &20:30:05  &1151  & 35   &N19W01  &  54  &  N16E08  &   1520  &   55 & 74   & 29\\
2002/07/16 &16:02:58  &1636  &325   &N23W07   & 58  &  N04W01  &   2467  &   235 & 86  & 346\\
2002/07/18 &08:06:08  &1099   &354   &N19W30  & 84  &  N18W12 &  1128   &  101 & 66   & 326\\
2002/07/26  &22:06:10  & 818  &172   &S19E26  & 96  & S25E16  &   804   &   41 & 60 &   149\\
2002/08/16 & 12:30:05  &1585  &121   &S14E20   &103 &  S15E06  &  1852   &   78 & 73  & 158\\
2002/09/05  &16:54:06  &1748  &114   &N09E28  & 47  &  S08E12  &  2770   &   56 & 75   & 123\\
2002/11/09 & 13:31:45  &1838  &233   &S12W29  & 46  &  S08W09  &  2770   &    137 & 78  & 228\\
2002/12/19 &22:06:05 &1092    &300  &N15W09   & 62  &  N06W01   &  2506   &   179  & 84 &  353\\

 \hline
\end{tabular}
}
\end{table*}

\end{article}

\end{document}